\documentclass[11pt]{article} 
\usepackage{hyperref} 
\pdfoutput=1

\begin{document} 
 
\title{Fast, large amplitude vibrations of compliant cylindrical shells carrying a fluid} 
 
\author{Pawel Zimoch, Eliott Tixier, Julia Hsu, Amos Winter and Anette Hosoi \\ 
\\\vspace{6pt} Department of Mechanical Engineering\\ Massachusetts Institute of Technology\\ Cambridge, MA 02139, USA} 
 
\maketitle

\begin{abstract} 
In this fluid dynamics video, we demonstrate the first three circumferential modes of fast, large amplitude vibrations of compliant cylindrical shells carrying a fluid. 
\end{abstract}

 \section{Details}
 We use shells made of vinylpolysiloxane (a silicone-based elastomer) with elastic moduli of $0.2$, $0.6$ and $1.0$ MPa, as well as shells made of latex. The shells range from $5$--$10$ mm in diameter, $5$--$50$ mm in length and approximately $0.1$--$0.2$ mm in thickness. We clamp one end of these shells onto a rigid nozzle and pass air through them at flow rates ranging from $0.2$--$2.5$ liters per second. When the flow rate of air in the shells exceeds a certain critical value, dependent on the dimensions and material properties of each shell, the shell becomes unstable and begins to vibrate. The mode of vibration corresponds to one of the circumferential normal modes of vibration of cantilevered cylindrical shells. Which mode is observed depends on the dimensions and material properties of the shell. We observed the first three modes. 
 
 In the first mode, commonly known as the ``garden hose mode," the shells oscillate side-to-side with the frequency of approximately $15$ Hz. 
 
 In the second mode, the surface of the shell bends inwards, obstructing the fluid flow and causing a large jump in the pressure drop across the nozzle. In this mode, the shell can vibrate with frequencies from $200$--$700$ Hz, depending on the volumetric flow rate of air. We observed that the frequency of oscillation is directly proportional to air flow rate. Additionally, when the shell vibrates in the second mode with average frequencies ranging from approximately $350$--$550$ Hz, the vibration is unstable and the oscillation frequency varies widely between periods. The second mode is the most robust and can be observed in the largest range of parameters in our shells. 
 
 In the third mode of vibration, the circumference of the free end of the shell is divided into three ``flaps" oscillating inwards and outwards. In this mode, the shells vibrate with frequency of approximately $600$--$1000$ Hz.
 
 The images were captured with a Phantom v5.2 high speed color camera. The images of the second and third mode were captured using a stroboscopic technique and the final video is a concatenation of frames taken from different oscillation periods. Each frame is slightly offset in phase, yielding a slow-motion effect.
 
 \section{Acknowledgments}
 
 We would like to express our gratitude to Felice Frankel for her help with cinematography, and to Prof. John Bush who kindly lent us the high speed camera.  We would also like to thank Prof. Pedro Reis and Dr. Arnaud Lazarus for sharing with us their data on elastic properties of vinylpolysiloxane elastomers.

\end{document}